\documentclass[a4paper,12pt]{article}
\begin{document}
\title{Mixing and irreversibility in classical mechanics
}
\author{V.M. Somsikov}
\date{\it{Laboratory of Physics of the geoheliocosmic relation,\\
Institute of Ionosphere, Almaty, 480020, Kazakhstan\\E-mail:
nes@kaznet.kz}} \maketitle

\begin{abstract}
The mechanism of irreversible dynamics in the systems with mixing
is analyzed. The procedure of splitting of system on equilibrium
subsystems and studying of dynamics of one of them under condition
of its interaction with other subsystems in the basis of the
approach to the analysis of dynamics of nonequilibrium systems is
used. The problem of "coarse-grain" of the phase space in this
method is eliminated. The formula, which expresses the entropy
through the work of forces between systems, is submitted. The
essential link between thermodynamics and classical mechanics was
found.
\end{abstract}.

\section{Introduction}
Irreversibility is an essential part of the second law of
thermodynamics in fundamental physics. According to this law there
is function $'S'$ named entropy, which can only grow for isolated
systems, achieving a maximum at a state of equilibrium. But this
is in contradiction with reversibility of the Newton equation and
potentiality of the fundamental forces [1, 2]. The great
importance of the irreversibility problem for fundamental physics
explains its big popularity among the physicists. The history of
its solution was very extensive and sometimes, dramatic.
Therefore, let us reference only few those works, which precisely
enough gives a clear picture of the state in the problem of
irreversibility [1, 3, 8, 17].

A first attempt to resolve this contradiction has been done by
Boltzmann. From the $H$-theorem it follows that many-body systems
should equilibrate. But for obtaining this result, Boltzmann had
used principles of probability. Therefore the contradiction was
not overcome. For overcoming this problem, it was suggested to try
to create an expanded formalism of open systems within the
framework of classical mechanics laws [9]. It turned out that such
formalism appears in the process of solving the problem of
irreversibility [12-15].

The substantiation of the mechanism of equilibration for
non-dissipative hard-disks system was based on the dependence of
force of disks interaction on their relative velocities, and on
the necessary condition for irreversibility [12]. The dependence
of the force of disks interaction on the velocities followed from
their equation of motion. The existence of irreversibility
condition followed from the general Liouville equation. But in
natural systems the forces of interacting of the elementary
particles are potential [2] and therefore the equation of motion
is reversible. Thus there is a key question on a way of a
substantiation of thermodynamics: is it possible to connect the
fact of the potentiality of interaction elements in the system
with irreversibility which in real systems exists? The search of
the answer on this question is the purpose of this work.

The investigation is based on the next method. A conservative
system of interacting elements, which is not in equilibrium, is
prepared. This system is then split into small subsystems that are
accepted as being in equilibrium. The subsystem dynamics under
condition of their interactions is analyzed on the basis of
classical mechanical laws.

The work was carried out in the following way. First of all, based
on the equation of motion for disks, we will show that the forces
between the selected subsystems, which we called the generalized
forces, are dependent on the velocities. This result has led us to
the key point that on the base of usual canonical Hamilton and
Liouville equations, the fundamental problem of irreversibility
cannot be solved. Instead we have used so-called generalized
Hamilton and Liouville equations [12]. Based on this, we discover
the condition of irreversible dynamics. This condition follows
from the generalized Liouville equation.

Then the dynamics of the systems constructed from potentially
interaction elements is analyzed. The equation for the energy
exchange between subsystems is obtained. Based on this equation we
answer the question of why and how the velocity dependence of
generalized forces between subsystems appears when the forces
between the elements are potential. Then the essential link
between thermodynamics and classical mechanics is analyzed. A
formula, which expresses the entropy through the work of
generalized forces, is obtained. This formula is determined by the
fact that the energy of subsystem interaction is transformed into
internal energy as a result of the work done by the generalized
forces in an irreversible way.

\section{Irreversibility in a hard-disks system}
The study of a hard-disk system is based on the equation of motion
for disks. This equation is deduced with help of the matrix of
pair collisions. In the complex plane this matrix is given [11]:

$S_{kj}=\left(\begin{array}{cc} a & -i b
\\ -i b & a\end{array} \right)$ (a),

where $a=d_{kj}\exp(i\vartheta_{kj})$; $b=\beta
\exp(i\vartheta_{kj})$; $d_{kj}=cos\vartheta_{kj}$;
$\beta=sin\vartheta_{kj}$; $i$ is the imaginary unit; $k$ and $j$
numbers of colliding disks; $d_{kj}$ is the impact parameter (IP),
determined by the distance between the centers of colliding disks
in a Cartesian plane system of coordinates with axes of $x$ and
$y$, in which the $k$-disk swoops on the lying $j$- disk along the
$x$ - axis. The scattering angle $\vartheta_{kj}$ varies from $0$
to $\pi$. In consequence of collision the transformation of disks
velocities can be presented in such form: $V_{kj}^{+}=S_{kj}
V_{kj}^{-}$ (a), where $V_{kj}^{-}$ and $V_{kj}^{+}$ - are
bivectors of velocities of $k$ and $j$ - disks before $(-)$, and
after $(+)$ collisions, correspondingly; $V_{kj}$=$\{V_k,V_j\}$,
${V}_j=V_{jx}+iV_{jy}$ - are complex velocities of the incident
disk and the disk - target with corresponding components to the
$x$- and $y$- axes. The collisions are considered to be central,
and friction is neglected. Masses and diameters of disks, $"d"$,
are accepted to be equal to $1$. Boundary conditions are given as
either periodical or in form of hard walls. From (a) we can obtain
equations for the change of velocities of colliding disks:
\begin{equation}
{\left(\begin{array}{c} {\delta V_k}\\ {\delta V_j}
\end{array} \right )
=\varphi_{kj} \left( \begin{array}{c} \Delta_{kj}^{-} \\
-\Delta_{kj}^{-}\end{array} \right )}.\label{eqn1}
\end{equation}
Here, $\Delta_{kj}=V_k-V_j$ are relative velocities, $\delta
V_k=V_k^{+}-V_k^{-}$, and $\delta V_j=V_j^{+}-V_j^{-}$ - are
changes of disks velocities in consequence of collisions,
$\varphi_{kj}=i\beta e^{i\vartheta_{kj}}$. \\That is, Eq. (1) can
be presented in the differential form:
\begin{equation}
\dot{V}_k=\Phi_{kj}\delta (\psi_{kj}(t))\Delta_{kj}\label{eqn2}
\end{equation}
where $\psi_{kj}=[1-|l_{kj}|]/|\Delta_{kj}|$;
$\delta(\psi_{kj})$-delta function; $l_{kj}(t)=z_{kj}^0+\int
\limits_{0}^{t}\Delta_{kj}{dt}$ are distances between centers of
colliding disks; $z_{kj}^0=z_k^0-z_j^0$, $z_k^0$ and $ z_j^0$ -
are initial values of disks coordinates;
$\Phi_{kj}=i(l_{kj}\Delta_{kj})/(|l_{kj}||\Delta_{kj}|)$.

The Eq. (2) determines a redistribution of kinetic energy between
the colliding disks. It is not a Newtonian equation because the
forces between the colliding disks depend on their relative
velocities. Hence, for the analysis of systems of disks it is
impossible to use the canonical Hamilton equation [4]. So, we get
the generalized Hamilton equation to be applied for studying the
subsystem dynamics [12]:

\begin{equation}
{\frac{\partial{H_p}}{\partial{r_k}}=-\dot{p_k}+F_{k}^p}\label{eqn3}
\end{equation}
\begin{equation}
{\frac{\partial{H_p}}{\partial{p_k}}={V_k}}\label{eqn4}
\end{equation}
These are the general Hamilton equations for the selected
$p$-subsystem. The external forces, which acted on $k$ disks
belong to the $p$-subsystem, presented in a right-hand side on Eq.
(3). These forces are not potential.

Using Eqs. (3,4), we can find the Liouville equation for
$p$-subsystem. For this purpose, let us take a generalized current
vector $J_p=(\dot{r_k},{\dot{p_k}})$ of the $p$-subsystem in a
phase space. From Eqs. (3,4), we find [12, 13]]:
\begin{equation}
{\frac{df_p}{dt}=-f_p\sum\limits_{k=1}^T
\frac{\partial}{\partial{p_k}}F_{k}^p} \label{eqn5}
\end{equation}

Eq. (5) is a Liouville equation for the $p$-subsystem. We can
rewrite it in differential form as: \\$\frac{df_p}{dt}+f_p
div_{\vec{p}}{\vec{F}}=0 $, where $\vec{p}=\{p_k\};
\vec{F}=\{F^{p}_k\}, k=1,2...T $.

The Liouville equation has the formal solution:
${f_p=const\cdot{\exp{[-\int\limits_{0}^{t}{(\sum\limits_{k=1}^T
\frac{\partial}{\partial{p_k}}F_{k}^p)}{dt}]}}}$.

The Eq. (5) is obtained from the common reasons. Therefore it is
suitable for any interaction forces of subsystems, as dissipative,
as non-dissipative. For a hard-disks system the energy dissipation
does not exist. There is only a redistribution of kinetic energy
between colliding disks. Thus, Eq. (5) is applicable to analyze
any open nonequilibrium systems, because it takes into account the
energy exchange between subsystems. Therefore it can be used also
for the explanation of irreversibility in a frames of the
classical mechanics laws.

The right side of Eq. (5), $f_p\sum\limits_{k=1}^T
\frac{\partial}{\partial{p_k}}F_{k}^p$, is a similar to the
integral of collisions. It can be obtained using the subsystems
motions equations. For a hard disks system it can be found with
the help of Eq. (2).

In the non-equilibrium system the right term of Eq. (5) is not
zero, because generalized forces are dependent on velocities.
Therefore the relative subsystems velocities are distinguish from
zero. This conclusion is in agreement with the fact, that
subsystems velocities in the non-equilibrium systems are non-zero
[5]. This can mean only, that when the system goes to equilibrium
state, the relative velocities of subsystems go to zero.

Let us consider the important interrelation between descriptions
of dynamics of separate subsystems and dynamics of system as a
whole. As the expression,
${\sum\limits_{p=1}^R{\sum\limits_{k=1}^T F_{k}^p =0}}$, is
carried out, the next equation for the full system Lagrangian,
$L_R$, will have a place:
\begin{equation}
{\frac{d}{dt}\frac{\partial{L_R}}{\partial{V_k}}-
\frac{\partial{L_R}}{\partial{r_k}}=0}\label{eqn6}
\end{equation}
and the appropriate Liouville equation:
${\frac{\partial{f_R}}{\partial{t}}+{V_k}\frac{\partial{f_R}}
{\partial{r_k}}+\dot{p_k}\frac{\partial{f_R}}{\partial{p_k}}=0}$.
The function, $f_R$, corresponds to the full system. The full
system is conservative. Therefore, we have: ${\sum\limits_{p=1}^R
divJ_p=0}$. This expression is equivalent to the next equality:
${\frac{d}{dt}(\sum\limits_{p=1}^{R}\ln{f_p})}=
\frac{d}{dt}(\ln{\prod\limits_{p=1}^{R}f_p})=
{(\prod\limits_{p=1}^{R}f_p)}^{-1}\frac{d}{dt}(\prod\limits_{p=1}^{R}{f_p})=0$.
So, $\prod\limits_{p=1}^R{f_p}=const$. In an equilibrium state we
have $\prod\limits_{p=1}^R{f_p}=f_R$. Because the equality
$\sum\limits_{p=1}^{R}F_p=0$ is fulfilled during all time, we have
that equality, $\prod\limits_{p=1}^R{f_p}=f_R$, is a motion
integral. It is in agreement with Liouville theorem about
conservation of phase space [4].

So, only in two cases the Liouville equation for the whole
non-equilibrium system is in agreement with the general Liouville
equation for selected subsystems: if the condition
$\int\limits_{0}^{t}{(\sum\limits_{k=1}^T\frac{\partial}
{\partial{p_k}}F_{k}^p)}dt\rightarrow{const}$ (c) is satisfied
when $t\rightarrow\infty$, or when,
${(\sum\limits_{k=1}^T}\frac{\partial}{\partial{p_k}}F_{k}^p)$, is
a periodic function of time. The first case corresponds to the
irreversible dynamics, and the second case corresponds to
reversible dynamics.

Because the generalized forces for a hard-disks system depended on
velocities, the irreversible dynamics is possible.

Dynamics of strongly rarefied systems of potentially interacting
elements is also described by the Eq. (2)[12]. Therefore for those
systems, irreversibility is possible as well.

Thus, the generalized Liouville equation allows describing
dynamics of nonequilibrium systems within the framework of laws of
classical mechanics. According to this equation the reversible and
irreversible dynamics have a place. Irreversibility is possible
only at presence of dependence of the generalized forces from
subsystems velocities. The dependence of the generalized forces on
velocities deletes the restriction on irreversibility,
superimposed by the Poincare's theorem about reversibility [17]
because this theorem is applicable to the systems with the
potential forces only.

We see that the dependence of generalized forces on velocities for
subsystems interaction is a necessary condition for the
irreversibility to occur. So, the question about irreversibility
for Newtonian systems is reduced to that about dependence of the
forces between subsystems on the velocity.

For a hard-disks system and for strongly rarefied system of
potentially interacting elements the presence of irreversibility
is predetermined by Eq. (2). In these systems the interaction
forces between the elements are depending on velocities. Therefore
it is clear that the generalized forces will depend on velocities
as well. But the forces between elements for Newtonian systems are
potential. Therefore it is necessary to answer the question: how
does velocity dependence of generalized force between subsystems
appear when forces between the elements are independent on
velocities. The answer on this question is a purpose of the next
part of the paper.

\section{The subsystems equation of motion }
Let us to analyze Newtonian systems. We take a system with energy:
$E_N=T_N+U_N=const$, where
$T_N=\frac{1}{2}\sum\limits_{i=1}^N{{v_i}^2}$ is a kinetic energy;
$U_N(r_{ij})$ is potential energy; $r_{ij}=r_i-r_j$ is the
distance between $i$ and $j$ elements; $N$ is the number of
elements. Masses of elements are accepted to $1$.

The Newton equation of motion for elements is:
\begin{equation}
{\dot{v}_i=-\sum\limits_{i=1,j\neq{i}}^N\frac{\partial}
{\partial{r_{ij}}}U}\label{eqn7}
\end{equation}

This equation in generally is nonlinearly. For two-body system by
transition in to the system of coordinates of the center of mass,
the nonlinearity can be eliminated, as kinetic energy of the
system's motion is excluded. In a result the system becomes
integrated. For the systems of three and more bodies the excluding
of the nonlinearity  in generally is impossible. Therefore they
are non-integrated. From here, by the way, the reason why the
nonequilibrium system should be splitting into equilibrium
subsystems, becomes clear. By such splitting we exclude
nonlinearity of dynamics inside subsystems . If the system is
equilibrium, that, as though we did not split  her into
subsystems, these subsystems will be motionless be relative each
other [5]. To emphasize the absence of energy of relative motion
of subsystems in the internal energy of equilibrium system, we
shall name internal energy of equilibrium system as a binding
energy.

Therefore, for the description of evolution of nonequilibrium
system, it is necessary to obtain an equation for energy exchange
between subsystems, which determines the generalized forces. For
this purpose we will do the following. In laboratory system of
coordinates we represent the total subsystem energy as the sum of
the kinetic energy of subsystem motion as the whole, $T_n^{tr}$,
the kinetic energy of its elements concerning the center of mass,
$\widetilde{T}_N^{ins}$, the potential energy of its elements
inside the subsystem, $\widetilde{U}_N^{ins}$, and finally it is
an energy of interaction with other subsystems.

The energy,
$E_N^{ins}=\widetilde{T}_N^{ins}+\widetilde{U}_N^{ins}$, is a
internal energy. The internal energy is the sum of kinetic energy
of relative motion of particles and their energy of potential
interaction. Its value is equal to full energy of a system minus
potential energy of its interaction with other systems and kinetic
energy of motion of a system as the whole. The internal energy,
$E_N^{ins}$, is determined by relative velocity of the particles
and distances between them. If a system is in equilibrium we will
call internal energy as binding energy.

The energy, $T_n^{tr}$, is determined by velocity of the center of
mass of system. It is depend on relative velocities of subsystems,
which depend on the orderliness of motion of particles. Therefore
the energy $T_n^{tr}$ is determine the degree of orderliness of
the system.

At absence of external forces, the energies, $T_N^{tr}$, and
$E_N^{ins}$, according to the law of preservation of a momentum of
all systems, are motion integrals.

Let us assume for simplification that the system is consist from
two interacting subsystems, both of them is in equilibrium. The
equations for the energy exchange for two subsystems have the
forms [14, 15]:

\begin{equation}
{LV_L\dot{V}_L+{\sum\limits_{j=i+1}^L}\sum\limits_{i=1}^{L-1}\{v_{ij}
[\frac{\dot{v}_{ij}}{L}+\frac{\partial{U}}{\partial{r_{ij}}}]\}=
-\sum\limits_{{j_K}=1}^K}\sum\limits_{{i_L}=1}^{L}v_{i_L}
\frac{\partial{U}}{\partial{r_{{i}_{L}{j}_{K}}}} \label{eqn8}
\end{equation}

\begin{equation}
{KV_K\dot{V}_K+{\sum\limits_{j=i+1}^K}\sum\limits_{i=L+1}^{K-1}\{v_{ij}
[\frac{\dot{v}_{ij}}{K}+\frac{\partial{U}}{\partial{r_{ij}}}]\}=
-\sum\limits_{{j_K}=1}^K}\sum\limits_{{i_L}=1}^{L}v_{j_K}
\frac{\partial{U}}{\partial{r_{{i}_{L}{j}_{K}}}} \label{eqn9}
\end{equation}

Here we take ${LV_L+KV_K=0}$, ${V_L}$ and ${V_K}$ are the
velocities of the center of mass for the subsystems; $L$ and $K$
are the number of elements in the subsystems; ${v_{ij}}$ are the
relative velocities between the $i$ and $j$ elements; ${L+K=N}$.
Masses of elements are accepted to $1$. The sub-indexes denote to
which subsystems some elements belong.

The first term in the left side Eqs. (8, 9) respectively expresses
the rate of change of kinetic energy for the subsystems,
${T^{tr}}$. The second term is related to transformation of
binding energy for the subsystems, ${E^{ins}}$.

The right side in the Eqs. (8, 9) determine the energy of
subsystems interaction. The interaction is a cause of the kinetic
energy transformation of the subsystem motion, ${T^{tr}}$, into
the binding energy.

The particles velocities can be write as a sum of velocity of the
center of mass of subsystem and velocity relation on it. So,
$v=V+\tilde{v}$. Then, we can write in case ${L=K}$ from the Eq.
(8):
\begin{equation}
{LV_l{[\dot{V}_L+\sum\limits_{{j_K}=1}^K}\sum\limits_{{i_L}=1}^{L}
\frac{\partial{U}}{\partial{r_{{i}_{L}{j}_{K}}}}] +
{\sum\limits_{j=i+1}^L}\sum\limits_{i=1}^{L-1}\{v_{ij}
[\frac{\dot{v}_{ij}}{L}+\frac{\partial{U}}{\partial{r_{ij}}}]\} =
-\sum\limits_{{j_K}=1}^K}\sum\limits_{{i_L}=1}^{L}
\frac{\partial{U}}{\partial{r_{{i}_{L}{j}_{K}}}}{\tilde{v}_{i_L}}
\label{eqn10}
\end{equation}

The first term in the left side Eqs. (10) respectively expresses
the rate of change of energy for the subsystem as a whole. The
second term is related to transformation of binding energy for the
subsystem. So the Eq. (10) determines the $L$ subsystem energy
change in interaction with the $K$ subsystem. This change is
determined by the general force, which depend on not only on the
coordinates but on the velocities as well.

Let us notice that in the laboratory system of coordinates, the internal energy of the system is equal to the sum of internals energies of subsystems. So the interaction of subsystems do not change they internal energies. But if we take the binding energy, which is determined concerning to the center of weights subsystems, all is differently. The binding energy will increase due to increasing of randomness of velocities vectors of the particles. This increase will occur due to the energy of relative motion of subsystems, which as a matter of fact is a measure of the order of system.

Let's compare the equation of Newton and the equation (7). The Newton equation can be treated, as the equation for potential forces. The work of these forces determines transformation of kinetic energy of elements to their potential energy. This transformation has a place at transition of system from one point of configuration space to another. Forces are determined by a gradient of potential energy of particles. Thus, the forces and the potential energy of particles are completely determined by coordinates. The work of potential forces on the closed contour is equal to zero. It is corresponds to reversible dynamics.

Let us consider now the equation (10). From it follows, that in nonequilibrium systems the kinetic energy of relative motion of subsystems is exist. It is connected with the orderliness of motion of particles of subsystems. Therefore it's determined by their function of distribution. As against potential forces, work of the generalized forces will transform kinetic energy of motion of a subsystem not only to the potential energy of a subsystem as the whole, but also into the binding energy. Because of such transformation of kinetic energy of motion of subsystems, the work of the generalized forces on the closed contour in configuration space differs from zero. Really, in the same point of the configuration space, at the same values of kinetic energy of motion of particles, the kinetic energy of motion of subsystems can be various because of a various degree of orderliness of motion of particles. Thus, in nonequilibrium systems there is a new type of a stream of energy. This energy is connected to relative motion of subsystems. It is determined by a degree of a deviation of velocities distribution function of particles from the equilibrium.

The eq. (10) is transformed into the Newton equation in three
cases: in equilibrium state; if subsystems can be taken as a hard
without internal degrees of freedom; if the subsystems are
conservative.

There is a question, why for the closed system the Newton equation for particles fairly, but, nevertheless, it does not determine evolution of system to equilibrium? It is because the work of the generalized forces determines transformation of kinetic energy of relative motion of subsystems into the binding energy. This transformation occurs in a result of approach of the velocity distribution of function of the subsystems elements to the equilibrium. Thus kinetic energy of relative motion of subsystems disappears, though total kinetic energy of particles of system can be saved. Really, in the system of coordinates of the center of mass it is possible to find decreasing of the orderliness of the particles velocities in a result of interaction of subsystems. It leads to reduction of relative velocities and energy of motion of subsystems, and increase of the binding energy. In result the system becomes equilibrium. All this process occurs due to disordering of the vectors of velocities of particles. But the Newton equation determines only the transition of kinetic energy of particles to the potential energy. So, the Newton equation "does not react" to such transformations of energy that do not change a ratio between kinetic and potential energy of the particles.

Irreversibility is caused by that the return of the binding energy to kinetic energy of a subsystem is impossible. Really, this return would be possible only under condition of spontaneous occurrence of the generalized forces inside an equilibrium subsystem. But their occurrence would mean broken of spherical symmetry of the distribution function of velocities of elements. It is also in contradiction with the law of preservation of a momentum.

Thus, the equation (10), as against of the Newton equation,
describes process of transformation of energy in the system,
caused not only by the transformation of the potential energy to
kinetic energy, but also change of the distribution function of
velocity of particles. This change of function of distribution is caused
by increase in a randomness of velocities of particles as a result of
mixing of the phase trajectories.

It is not difficult to generalize the equation (7) on a case when forces of interactions of particles inside subsystems will be much more than forces of interaction between particles of different subsystems. This case corresponds, or to different types of potential interaction of particles of different subsystems, or big enough distance between subsystems. If interaction forces of particles inside subsystems are infinite, the right part (7) is equal to zero. Then the change of the binding energy of subsystems will be zero and the equation (7) will transformed to the usual equation of Newton for interacting hard body.

Having excluded from the equation (7) of the potential energy, we
shall obtain the equation of motion of a hard disks. It means,
that both, in systems of hard disks, and in systems of potentially
interacting elements, the nature of irreversibility is
identically. The irreversibility is determined by transformation
of energy of relative motion of subsystems to their binding energy
as a result of amplification of the disorder.

In equilibrium the relative
velocities of subsystems and the energy flow between them are
equal to zero for any splitting [5]. The equilibrium state is
stable. From the physical point of view the stability of an
equilibrium state for mixed systems is caused by aspiration to
zero of the generalized forces arising at a deviation of system
from equilibrium [10, 13]. Hence, the system, having come to
equilibrium, never leaves this state.

\section{Thermodynamics and classical mechanics}
With the help of the equations (8-10) it is possible to link classical mechanics with thermodynamics. Really, the right side of these equations determines an exchange of energy between subsystems as a result of their interaction. The first term of the left side of each equation determines change of energy of motion of the subsystem as the whole. In thermodynamics it corresponds to mechanical work done with subsystem by external influence. The second term of the left side corresponds to increase of the binding energy of a subsystem. In thermodynamics this term corresponds to change of thermal energy of system.

Let us now consider the essence link between thermodynamics and
classical mechanics. It is easy to see the analogy between the
Eqs. (8-10) and the basic equation of thermodynamics [10]:

\begin{equation}
{dE=dQ-{PdY}} \label{eqn11}
\end{equation}

Here, according to common terminology, $E$ is internal energy of a
subsystem; $Q$ is thermal energy; $P$ is pressure; $Y$ is volume.

The energy change of the selected subsystem is due to the work
made by external forces. Therefore, the change in full energy of a
subsystem corresponds to $dE$.

The change of kinetic energy of motion of a subsystem as the
whole, $dT^{tr}$, corresponds to the term ${PdY}$. Really,
${dT^{tr}=VdV=V\dot{V}dt=\dot{V}dr=PdY}$

Let us determine, what term in Eq. (11) corresponds to the change
of the binding energy in a subsystem. As follows from virial
theorem [6], if the potential energy is a homogeneous function of
second order of the radiuses-vectors, then
${\bar{E}^{ins}=2\bar{\tilde{T}}^{ins}=2\bar{\tilde{U}}^{ins}}$.
The line denotes the time average. Earlier we obtained that the
binding energy, ${E^{ins}}$, increases due to contribution of
energy, ${T^{tr}}$. But the opposite process is impossible.
Therefore the change of the term $Q$ in the Eq. (11)  corresponds
to the change of the binding energy ${E^{ins}}$.

Let us consider the system near to equilibrium. If the subsystem
consist of ${N_m}$ elements, the average energy of each element
becomes, ${\bar{E}^{ins}={E}^{ins}/N_m=\kappa{T}_0^{ins}}$. Now
let the binding energy increases with ${dQ}$. According to the
virial theorem, keeping the terms of the first order, we have:

${dQ\approx{T}_0^{ins}[d{E}^{ins}/{T}_0^{ins}]
={T}_0^{ins}[{dv}/{v_0}]}$, where ${v_0}$ is the average velocity
of an element, and ${dv}$ is its change. For subsystems in
equilibrium, we have ${dv/v_0\sim{{d\Gamma_m}/{\Gamma_m}}}$, where
${\Gamma_m}$ is the phase volume of a subsystem, ${d\Gamma_m}$
will increase due to increasing of the subsystem energy on the
value, ${dQ}$. By keeping the terms of the first order we get:
${dQ\approx{T}_0^{ins}d\Gamma_m/\Gamma_m={{T}_0^{ins}}d\ln{\Gamma_m}}$.
By definition ${d\ln{\Gamma_m}=dS^{ins}}$, where ${S^{ins}}$ is a
subsystem entropy [5, 10]. So, near equilibrium we have
${dQ\approx{T}_0^{ins}dS^{ins}}$.

\section{Relation between entropy and generalized forces}

Let us consider the relation of the generalized field of forces
with entropy. According to the results obtained here, and also in
agreement with [5] the equilibrium state
of the system is characterized by absence of energy of relative
motion, ${T_m^{tr}}$, of subsystems. I.e. energy ${T^{tr}}$, as a
result of the work done by the generalized forces, will be
redistributed between the subsystems into the binding energy. This
causes an increase of entropy. When the relative velocities of the
subsystems go to zero, the system goes to equilibrium. So, the
entropy increasing, $\Delta{S}$, can be determined as follows :

 \begin{equation}
{{\Delta{S}}={\sum\limits_{l=1}^R{\{{m_l}
\sum\limits_{k=1}^{m_l}\int{\sum\limits_s{{\frac{{F_{ks}}^{m_l}v_k}{E^{m_l}}}}{dt}}\}}}}\label{eqn12}
\end{equation}

Here ${E^{m_l}}$ is the kinetic energy of subsystem; ${m_l}$ is
the number elements in subsystem ${"l"}$; ${R}$ is the number of
subsystems; ${s}$ is number of the external disks which collided
with internal disk ${k}$. The integral is determining the work of
the force ${F_{ks}^{m_l}}$ during the relaxation to equilibrium.
It is corresponds to phenomenological formula Clauses for entropy
[5, 10]. So, Eq. (12) deduced entropy through
the generalized force. Therefore we can use this equation for
analyzing different kind of entropy [3, 16,].

Inherently Eq. (12) corresponds to the formula for entropy:
${S=\sum\limits_{l=1}^R{S_l(E_l^{ins}+T_l^{tr})}}$, (see, [5]). In fact, if ${E_l^{ins}\gg{T_l^{tr}}}$,
then
${dS=\sum\limits_{l=1}^R\frac{\partial{S_l}}{\partial{T_l^{tr}}}{dT_l^{tr}}}$
which corresponds to Eq.(12).

Thus, the Eq. (12) connects the force acting on the system with entropy.
I.e. this formula established the connection between
parameters of classical mechanics and thermodynamic parameters. It
determines a measure of a deviation of system from equilibrium.
Moreover, it opens essence of interrelation of Boltzmann entropy
definition, based on a measure of the chaos, and Clauses entropy,
based on the energy measure.

\section{Discussion}
From the time of statement of a problem of irreversibility   passed about 150 years. Up to now it was reduced to the problem of coarse-grain of the phase space. All attempts to solving this problem within the framework of classical mechanics were unsuccessful as they encountered Poincare's theorem of reversibility. This theorem based on a strict of canonical formalism of Hamilton. All this, apparently, deprives us any hopes for the successful solution of a problem of irreversibility within the framework of classical mechanics [17]. Is it so?

Already at the solution of a problem of three bodies there were
doubts concerning completeness of known methods of classical
mechanics. Difficulty of its solution, anyhow, is connected with a
problem of the description of the nonlinear
process of an exchange of energy between bodies. The same problem
forced scientists to be limited by researches only
equilibrium systems when it is possible to neglect an exchange
of energy between subsystems.
In kinetic physics where energy flows,
substances, etc. play a basic role, the
phenomenological formulas basing on statistical laws were used for their calculation [7]. But the exchange of energy between
particles and subsystems determines process of an establishment of
equilibrium. From here we also came to the conclusion that it is
impossible to solve a problem of irreversibility if not to find a
way of the description of process of an exchange of energy inside
systems within the framework of classical mechanics. It dictates
desire to execute such expansion of a Hamilton formalism which will
allow to describe dynamics of open systems.

Creation of the generalized formalism and studying of the
mechanism of irreversibility was carried out by us in parallel [11-15]. With
this purpose a conservative system of interacting elements, which
is not in equilibrium, is prepared. This system is then split into
small subsystems that are accepted as being in equilibrium. The
subsystem dynamics under condition of their interactions is
analyzed on the basis of classical mechanical laws.

First of all the dynamics of hard-disk systems is analyzed. Here,
based on the motion equation for disks [11], we showed that the
non-dissipative forces between the selected subsystems are dependent on the
velocities. This result has led us to the key point that on the
base of usual canonical Liouville equation, the problem of
irreversibility cannot be solved. Instead we have to use so-called
generalized Liouville equation [12, 13]. We showed that the
irreversible dynamics really does follow from the generalized
Liouville equation when forces between subsystems depend on
velocities.

But studying of dynamics of disks was preliminary step on a way to
understanding of the nature of irreversibility. Really, dynamics
of disks is determined by non-Newtonian forces. In a real
systems fundamental forces is potential [2]. Therefore these
results required generalization on a systems of potentially
interacting elements.

The results of disks research have shown, that the question on
irreversibility is reduced to a question on dependence
of the forces of interaction of subsystems on velocities. This statement
is in agreement with results, which was obtained in earlier works
by a similar method: splitting the system into equilibrium
subsystems and studying its dynamics near equilibrium state [5]. The macroscopically motion (it is similar to the subsystems
motion in our case) was considered. By using of the method of
Lagrange multiplier it has been strictly proved that equilibrium
is possible when, ${T^{tr}=0}$. Therefore it was necessary to
find out, whether irreversible transformation of kinetic
energy of relative motions of subsystems of potentially
interacting elements is possibly. The equation of Newton for this purpose did
not convenient. Really, it describes only such transformations of
energy, which are connected to reversible transition of kinetic
energy in potential. On the other hand, the equation of Newton is
correct for any systems. It has led us to suggestion, that the Newton equation is not responsible for transformation of the energy, connected with work of non-potential forces. Became clear from here why
the redistribution of energy caused by reduction of
relative velocities of subsystems as a result of increase of a
randomness of velocities of particles, is not described by
canonical Liouville equation and why the
coarse-grain is impossible. We came to the conclusion, that it is necessary to find a way of the description of process of transformation for two types of energy between systems. For this purpose analytical expression for an energy exchange between subsystems from the law of conservation of energy of system of potentially interacting elements has been obtained. From it expression followed, that the energy of relative motion of subsystems would be transformed to the binding energy as a result of disorder increasing of the vectors of velocities of particles. As a result, the relative velocities of subsystems will decrease. It has allowed offering the following mechanism of irreversible dynamics.

The nonequilibrium system is characterized by the greater
orderliness of elements motions. It causes existence of relative
motion of the equilibrium subsystems. As a result of
mixing, the disorder of vectors of
velocities of the particles is increase. It will lead to decrease of the
subsystems velocities and increasing of the binding energy. The
given process is irreversible because impossibly of increasing
relative velocities of motion due to the binding energy. This conclusion is
follows from the law of preservation of a subsystems momentum. As
a result the kinetic energy of relative motion of subsystems will
completely be transformed to the binding energy and system
equilibrates.

The offered scenario of irreversibility cannot be obtained on the
basis of the canonical equations of classical mechanics, as these
equations do not describe work of non-potential forces, that reducing
orderliness of motion of particles. Though mixing property
underlies in the offered scenario of irreversibility, the
coarse-grain problem of the phase space in it does not exist.

This scenario of irreversibility leads to a substantiation of
thermodynamics. Really, the first thermodynamics law follows from the equations (8-10) which describe
transformation energy of interacting subsystems. These
equations are determined by presence of two qualitatively various
types of energy: the binding energy and kinetic energy of relative
motion of a subsystem as the whole. Irreversible transition of
energy of motion of a subsystem to their binding energy
determines the contents of the second law of thermodynamics. I.e.
the energy of the motion of subsystems as a result of amplification of
chaos goes on increase of the entropy.

The further development of researches of irreversible dynamics on
the basis of the offered approach represents significant interest.
These researches will help explain connection between
thermodynamic laws and classical mechanics. They are perspective
also from the point of view of creation of the expanded formalism
of the classical mechanics, which necessary for investigation of
the open systems.

\smallskip

\end{document}